\documentclass{emulateapj}

\begin{document}

\title{The Far-Infrared Luminosity Function from GOODS-N: Constraining the Evolution of Infrared Galaxies for  $\lowercase{z}  \leq 1$}


\author{Minh T. Huynh}
\affil{Spitzer Science Center, MC220-6, California Institute of Technology,
Pasadena CA 91125, USA}
\email{mhuynh@ipac.caltech.edu}

\author{David T. Frayer}
\affil{Spitzer Science Center, MC220-6, California Institute of Technology,
Pasadena CA 91125, USA}

\author{Bahram Mobasher}
\affil{Space Telescope Science Institute, 3700 San Martin Drive, Baltimore, MD 21218, USA}

\author{Mark Dickinson}
\affil{National Optical Astronomy Observatory, P.O. Box 26732, Tucson, AZ 85726, USA}

\author{Ranga-Ram Chary }
\affil{Spitzer Science Center, MC220-6, California Institute of Technology,
Pasadena CA 91125, USA}

\author{Glenn Morrison\altaffilmark{1}}
\affil{Institute for Astronomy, University of Hawaii, Honolulu, HI, 96822, USA}

\altaffiltext{1}{Also Canada-France-Hawaii Telescope, Kamuela, HI 96743 USA}

\begin{abstract}
We present the IR luminosity function derived from ultra-deep 70$\mu$m imaging of the GOODS-North field. The 70 $\mu$m observations are longward of the PAH and silicate features which complicate work in the MIR. We derive far-infrared luminosities for the 143 sources with $S_{70}> 2$ mJy  (S/N $> 3 \sigma$). The majority (81\%) of the sources have spectroscopic redshifts, and photometric redshifts are calculated for the remainder. The IR luminosity function at four redshifts ($z \sim$ 0.28, 0.48, 0.78, and 0.97) is derived and compared to the local one. There is considerable degeneracy between luminosity and density evolution. If the evolving luminosity function is described as $\rho(L, z) = (1 + z)^q \rho(L/(1 + z)^p, 0)$, we find $q = -2.19p + 6.09$. In the case of pure luminosity evolution, we find a best fit of $p = 2.78^{+0.34}_{-0.32}$. This is consistent with the results from 24$\mu$m and 1.4GHz studies. Our results confirm the emerging picture of  strong evolution in LIRGs and ULIRGs at $0.4 < z  < 1.1$, but we find no evidence of significant evolution in the sub-LIRG ($L < 10^{11} L_{\odot}$) population for $z < 0.4$.

\end{abstract}


\section{Introduction}

Deep mid-infrared surveys are revealing a population of mid and far-infrared luminous galaxies out to $z \sim 3$. These luminous (LIRGs, $10^{11} L_\odot <  L_{\rm IR} \equiv L_{8-1000\mu{\rm m}} < 10^{12} L_\odot$) and ultraluminous (ULIRGs, $L_{\rm IR} > 10^{12} L_\odot$) infrared galaxies are relatively rare in the local universe, but become increasingly important at high redshift, where dust enshrouded starbursts dominate the total cosmic star formation rate (e.g. \citealp{chary2001}, \citealp{blain2002}). 

The {\sl Infrared Space Observatory} (ISO) showed that infrared luminous starbursts were much more numerous at $z \sim 1$ than at the present time \citep{franceschini2001, elbaz2002}. The ISO results were expanded upon by deep surveys at 24 $\mu$m with the Multiband Imaging Photometer (MIPS) on the {\sl Spitzer Space Telescope} (e.g. \citealp{chary2004}, \citealp{papovich2004}). Using the excellent ancillary data in the Great Observatories Origins Deep Survey (GOODS) South and North fields, 15 $\mu$m and total infrared luminosity functions were derived from thousands of 24 $\mu$m sources \citep{lefloch2005, perez2005}. Strong evolution of the IR population was found and the IR luminosity function evolves as $(1 + z)^4$ for $z \lesssim 1$ \citep{lefloch2005, perez2005}.

The 24 $\mu$m results are dependent on the set of SED templates used to extrapolate the 24 $\mu$m flux densities to 15 $\mu$m and total infrared luminosities. Furthermore, significant variations in the bolometric correction are expected as strong PAH and silicate emission and absorption features are redshifted into the 24 $\mu$m band. Observations with the 70 $\mu$m band of MIPS are closer to the peak in FIR emission and are not affected by PAH or silicate features for $z \lesssim 3$. They should therefore provide more robust estimates of the far-infrared (FIR) luminosities.

Studies by ISO in the FIR regime have been limited in sensitivity ($S_{90 \mu{\rm m}} \gtrsim 100$ mJy, $S_{170\mu{\rm m}} > 200$ mJy) and redshift completeness \citep{serjeant2004, takeuchi2006}.
\cite{frayer2006} derived a FIR luminosity function (LF) for the Extragalactic First Look Survey (xFLS) from Spitzer 70 $\mu$m data, but this survey had incomplete redshift information at faint fluxes, and it was limited to $z < 0.3$ and bright ($S_{70\mu{\rm m}} \gtrsim 50$ mJy) sources.  In this paper we present the infrared luminosity function up to redshift 1 from the ultra-deep 70 $\mu$m survey of GOODS-N. 

We assume a Hubble constant of $71\,{\rm km}\,{\rm s}^{-1}{\rm Mpc}^{-1}$, and a standard $\Lambda$-CDM cosmology with $\Omega_{\rm M}=0.27$ and $\Omega_{\rm \Lambda}=0.73$ throughout this paper. We define the IR flux as the integrated flux over the wavelength range 8 to 1000 $\mu$m. 

\section{The Data}
\subsection{Ultra-deep 70 $\mu$m Imaging}

The GOODS-N field is centered on the Hubble Deep Field North at 12h36m55s, +62$^\circ$14m15s.
The MIPS 70$\,\mu$m observations of GOODS-N were carried out during Cycle 1 
({\sl Spitzer\/} program ID 3325, \citep{frayer2006b} and Cycle 3 (January 2006) for the Far Infrared Deep Extragalactic Legacy project (FIDEL, Spitzer PID:30948, PI: Dickinson). Together these data map a region 10\arcmin $\times$ 18\arcmin\,  to a depth of $10.6\,$ksec. 

The raw data were processed off-line using the Germanium Reprocessing Tools (GeRT), following the techniques described  in \cite{frayer2006b}.
We have cataloged 143 sources (over $\sim$$185\,{\rm arcmin}^2$) with $S_{70}\,{\gtrsim}\,2.0\,$mJy
(S/N$\,{>}\,3\sigma$) in GOODS-N. The 70$\,\mu$m images have a beam size of 18\farcs5 FWHM, and in the presence of Gaussian noise the 1$\sigma$ positional error of sources is of the order
$\frac{0.5\,\theta_{\rm FWHM}}{{\rm S/N}}$, i.e. 3\arcsec\  for the faintest sources. 

\subsection{Redshifts}

All 70 micron sources were matched to 24 micron and IRAC sources to obtain good positions. The best Spitzer position was then used to search for optical redshifts. About 7\% of the 70 micron sources have more than one 24 micron source within the 70 micron beam, and these were deblended individually (e.g. \citealp{huynh2007}). Spectroscopic redshifts are available for 116 of the 143 objects (\citealp{cohen2000}; \citealp{wirth2004}; Stern et al. in prep).

Photometric redshifts were derived for 141 of the 143 sources with the
extensive photometry available: ACS {\sl HST} \citep{giavalisco2004}, U- (NOAO), BVRIz-
(Subaru-SupremeCam) and JK- (NOAO/KittPeak-Flamingo) imaging. 
The photometric redshifts were calculated using
the $\chi2$ minimization technique as explained in \cite{mobasher2006}.
We have photometric redshifts for 26/27 sources that don't have a spectroscopic redshift and we therefore have redshift information for 142/143 sources. 

We quantified the reliability of the photometric redshifts by examining the fractional error, $\Delta \equiv 
(z_{\rm phot} - z_{\rm spec} / (1 + z_{\rm spec})$. For all 115 70 $\mu$m sources with both photometric and spectroscopic redshifts, we found the median fractional error, $\Delta$, is $0.012 \pm 0.20$. Assuming the 6 cases where the fractional error is greater than 0.2 are outliers, the success rate of the photometric redshift method is 95\%. Removing the 6 outliers gives a median fractional error of $0.0014 \pm 0.05$. We therefore conclude that the photometric redshifts are statistically reliable. 

The 70 micron sources have a median redshift of 0.64 (see Figure 1). The majority (79\%) of sources lie at  $z < 1$, as expected for the survey sensitivity and steep k-correction that is present at 70 micron. 

\section{Infrared Luminosities}

Many authors argue that the MIR is a good indicator of the bolometric IR luminosity for  normal and IR luminous galaxies (e.g. Chary and Elbaz 2001). Based on this, several authors have developed 
sets of galaxy templates that can be used to estimate the total infrared luminosity (\citealp{chary2001}; \citealp{dh02}; \citealp{lagache2003}).

We use the luminosity dependent SED templates based on local galaxies from Chary and Elbaz (2001) to determine the IR luminosities of the 70 $\mu$m galaxies. However it is not clear whether local templates can accurately reproduce the MIR SED of distant galaxies because PAH and silicate absorption features are dependent on complex dust physics, including the intensity of the radiation field, the metallicity of the ISM, and the distribution of grain sizes. For this reason we determine the IR luminosities of the 70 $\mu$m galaxies by fitting templates to the observed 70 $\mu$m flux density only, which is longward of the PAH and silicate features. 

The IR luminosities as a function of redshift are shown in Figure 1. Most of the sources below redshift $z = 1$ have LIRG-like luminosities. The higher redshift sources are luminous ULIRGs with possibly an embedded AGN.

The estimated accuracy of the IR luminosity, from the 70 $\mu$m flux density calibration and PSF fitting errors alone, is 9\%.  However, the luminosities derived are dependent on the SEDs used. The adopted template SEDs do not reflect the full range of SEDs observed in galaxies, and thus are the main source of systematic errors. For example, the total IR luminosity derived from the MIR regime can vary by a factor of 5 for local galaxies \citep{dale2005}. We are working longward of the PAHs and silicate features which affected previous work based on the MIR, but, on the other hand, the restframe wavelengths probed at 70 $\mu$m is affected by dust temperatures and emissivity. 

To test the consistency of our derived IR luminosities and the application of the adopted SEDs, we use the well known FIR-radio correlation. The deep radio image of GOODS-N (5 $\mu$Jy rms at 1.4 GHz, Morrison et al. in preparation) detects 120/143 (84\%) of the 70 $\mu$m sources at 3$\sigma$ or above. 
The FIR-radio correlation, $q = \log ({\rm FIR}/S_{\rm 1.4 GHz})$, where `FIR' here refers to the flux between 40 and 120 $\mu$m (e.g.~\citealp{yun2001}), has an observed
local value of $q\,{=}\,2.34\pm0.3$ \citep{yun2001}. Adopting an average factor of 2.0 between IR and FIR (e.g. \citealp{dh02}) and a radio spectral index of $\alpha = -0.8$\footnote{$S_\nu \propto \nu^\alpha$.}, we find $q\,{=}\,2.2\pm0.2$ for the radio detected sources. Including the 24 $\mu$m data in the fits to the SEDs gave a slightly larger dispersion in $q$. This suggests that the IR luminosities as estimated from the 70 $\mu$m data alone are reasonable.
 
\section{Infrared Luminosity Functions}

In this Section we explore the evolution of the IR luminosity function between redshifts 0 and 1.

\subsection{Methodology}

The luminosity functions were derived for 4 redshift bins, $0.2 < z < 0.4$, $0.4 < z < 0.6$, $0.6 < z < 0.9$, and $0.9 < z < 1.1$ using the usual $1/V_{\rm max}$ method (Schmidt 1968). These redshift bins were made wide enough so that there is a reasonable number of sources for calculating the luminosity function. The bins have median redshifts of 0.28, 0.48, 0.78, and 0.97, so a moderate range in redshift is explored. The comoving volume for each source is $V_{\rm max} = V_{z_{\rm max}} - V_{z_{\rm min}} $, where $z_{\rm min}$ is the lower limit of the redshift bin, and $z_{\rm max}$ is the maximum redshift at which the source would be included in the catalog, given the limiting 3$\sigma$ limit, or the maximum redshift of the bin. 

As mentioned in Section 2.2, we have almost complete redshift information on the 70 $\mu$m sample. A correction factor for each individual source was computed to correct for source detectability across the full image and flux boosting (i.e. the over-estimation of the flux densities of low SN sources). This correction was calculated using the Monte Carlo approach described by \cite{chary2004} and it is the same correction applied to the source counts \citep{frayer2006b}. 

\subsection{Results and Discussion}

The luminosity functions were derived from the restframe IR $\mu$m luminosities. In Figure 2 we plot the luminosity functions for the redshift bins explored, and the data is summarized in Table 1.  The local IR luminosity function from IRAS sources \citep{sanders2003} is plotted for comparison.

For each luminosity bin the uncertainties, $\sigma_{\rho}$, were estimated using the Poisson statistics on the number of sources, so  $\sigma_{\rho} = \left(\sum{\frac{1}{V_{\rm max}^2}}\right)^{1/2}$. Monte Carlo simulations were also performed to disentangle the uncertainties in the derivation of the luminosity function due to photometric errors. Each source was randomly given an IR luminosity within the uncertainty estimates and the luminosity function was re-calculated. We find this adds between 0.03 to  0.09 dex to the luminosity function uncertainty, depending on the bin, but the Poisson statistics dominate the uncertainties. 

We do not find any significant evolution in the sub-LIRG population ($L < 10^{11} L_{\odot}$) for the lowest redshift bin ($z < 0.4$) (Figure 2).  The high redshift LFs show evidence for strong evolution of LIRGs and ULIRGs at $z > 0.4$.

The IR LFs derived here are consistent with that derived from 24 micron \citep{lefloch2005} for the overlapping luminosity bins at $z < 1$. This implies that, on average, similar bolometric luminosities are derived from 24 and 70 micron for moderate luminosity ($L < 10^{11.8} L_{\odot}$) and moderate redshift sources ($z < 1$) sources. We can not say if this is the case for high luminosities and high redshifts (e.g. \citealp{chapman2005, pope2006}), as those sources are rare in the 70 micron data. Recent 70 micron stacking analysis of galaxies at $z \sim 2$ show that 24 micron observations at high redshift  over-estimate LIR in comparison to 70 micron and other LIR indicators \citep{daddi2007, papovich2007}. 

To explore the evolution of IR sources we use the analytical form of the local luminosity function (LF) from \cite{sanders2003} that comprises of a double power law: 
$\rho \propto L^{-0.6}$  for $\log (L/L_{\odot}) < 10.5 $, 
$\rho \propto L^{-2.2}$  for $\log (L/L_{\odot}) > 10.5 $.
We assume that the evolving luminosity function can be described by $\rho (L, z) = g(z) \rho[L / f(z), 0]$. In this sense $g(z)$ and $f(z)$ describe the density and luminosity evolution of the LF, respectively. The commonly used form of evolution is to assume $f(z) = (1 + z)^p$ and $g(z) = (1 + z )^q$ (e.g. \citealp{condon1984}; \citealp{haarsma2000}). 

Using $\chi^2$ minimization, we examine the best fit to the evolution of the IR LF. There is a well known degeneracy between density and luminosity evolution. We find that the best fit evolution parameters follow the relation $q = -2.19p + 6.09$. In the case of pure luminosity evolution ($q = 0$), we find $p = 2.78^{+0.34}_{-0.32}$. 

These evolution constraints are broadly consistent with 24 $\mu$m studies which found $p = 3.2^{+0.7}_{-0.2}$ and $q = 0.7^{+0.2}_{-0.6}$ for the infrared luminosity function \citep{lefloch2005}. Our results are also in good agreement with previous studies of IR sources (e.g. \citealp{franceschini2001}).  

\cite{hopkins2004} combined star formation rate data with faint radio source counts to find  $p = 2.7 \pm 0.6$ and $q = 0.15 \pm 0.60$. If only pure luminosity evolution of radio sources is considered then $p = 2.5 \pm 0.5$  \citep{Seymour2004} or $p = 2.7$ \citep{huynh2005}. So our results are consistent with constraints on the evolution of the star forming population from deep radio surveys, indicating that the radio sources overlap with the ultra-deep 70 $\mu$m population, as expected.

The constraints on the evolution of the IR LF can be used to determine the cosmic star formation rate (SFR) density. Using the calibration from \cite{kennicutt1998} and integrating over galaxies with $8.5 < \log(L/L_\odot) < 12.5$, we find the SFR density at $z = 1$ is $0.15^{+0.04}_{-0.03}$ $M_\odot$ yr$^{-1}$ Mpc$^{-3}$ for the best fit pure luminosity evolution case. Here the uncertainties in SFR density do not include the systematics in the FIR/SFR calibration, which add about 0.3 dex to the absolute uncertainty. The SFR density derived here is lower than that estimated by the evolutionary models of \cite{chary2001} by about a factor of 1.7, but it is consistent with extinction corrected optical measures (e.g. \citealp{kewley2004}) and 24 $\mu$m results \citep{lefloch2005}.

The AGN in our sample can be identified using the deep 2 Ms X-Ray observations of GOODS-N \citep{alexander2003}. Sources are classed as X-Ray AGN from X-ray band ratios, X-ray luminosity, and X-ray-to-optical flux ratios \citep{alexander2003,bauer2004}. At redshifts $z < 0.6$ we find only 7\% of the 70 $\mu$m sources are X-Ray AGN, but this fraction increases to 27\% for the $0.9 < z < 1.1$ redshift bin. The highest redshift LF in Figure 2 is contaminated by X-Ray AGN but this does not significantly affect the evolution derived in this work. 

\section{Concluding Remarks}

Based on ultra-deep 70 $\mu$m observations of GOODS-N, and the spectroscopic and photometric redshifts available of galaxies in this well-studied field, we have derived luminosity functions for $z = 0.3$ to $z = 1.1$.  We find strong evolution in galaxies with $L_{\rm IR} > 10^{11} L_{\odot}$ at redshifts $z > 0.4$. Assuming pure luminosity evolution of the form $(1 + z)^p$, we find $p = 2.78^{+0.34}_{-0.32}$. This confirms the strong evolution in LIRGs and ULIRGs between redshift 0 and 1 that has been seen in previous work.  The depth of the 70 $\mu$m data allows us to probe sub-LIRG luminosities, and we find little evolution in this population for $z \lesssim 0.4$. In the case of pure luminosity evolution, we find the star formation rate density at $z = 1$ is $0.15^{+0.04}_{-0.03}$ $M_\odot$ yr$^{-1}$  Mpc$^{-3}$.

This is the first result from an ultra-deep FIR survey that reaches $z \sim 1$.  However, we are limited by poor  statistics - the number of bins available for the LF at each redshift is limited by the small number of cataloged sources.  The area covered is only 10\arcmin $\times$ 18\arcmin\ so these results are also affected by cosmic variance.

The Far Infrared Deep Extragalactic Legacy (FIDEL) Spitzer legacy project, currently underway, will  cover the extended Chandra Deep Field South and the extended Groth Strip at 70 $\mu$m with similar depths to the GOODS-N. The total area covered will be about 9 times that used in this work, and the FIDEL project will detect over 1000 LIRGs at moderate redshift. So in the near future, large ultra-deep FIR surveys such as FIDEL will enable even more detailed studies of the FIR luminosity function and the evolution of infrared galaxies. 

\acknowledgements 

This work is based on observations made with the {\sl Spitzer Space Telescope}, which is operated by the Jet Propulsion Laboratory, California Institute of Technology under a
contract with NASA.  Support for this work was provided by NASA through an
award issued by JPL/Caltech. 

\bibliographystyle{aj}
\bibliography{refs}


\begin{table}[bt]
\centering
\begin{minipage}{\textwidth}
\caption {The IR luminosity functions derived from the $1/V_{\rm max}$ analysis.}
\begin{tabular}{lllccc}  \hline
log($L_{\rm IR}/L_{\odot}$) & log($L_{\rm IR}/L_{\odot}$) & log($L_{\rm IR}/L_{\odot}$) & N & $\rho ({\rm Mpc}^{-3} {\rm logL}^{-1})$\footnote{The errors quoted are  Poisson uncertainties.} & Monte Carlo \\
low & high & median & & & uncertainty\footnote{This is the additional uncertainty to be added to the LF, calculated from Monte Carlo simulations of the uncertainty in the IR luminosity of each source. See Section 4.2.} (dex)\\ \hline
\multicolumn{6}{c}{$0.2 < z < 0.4$} \\
10.20 & 10.50 & 10.44 & 7 & $3.36 \pm 1.58 \times 10^{-3}$  & 0.04 \\
10.50 & 10.70 & 10.63 & 7 & $2.44 \pm 0.94 \times 10^{-3}$ & 0.06 \\ \hline
\multicolumn{6}{c}{$0.4 < z < 0.6$} \\
10.60 & 10.90 & 10.81 & 9 & $1.77 \pm 0.65 \times 10^{-3}$ & 0.07 \\
10.90 & 11.10 & 11.00 & 11 & $1.68 \pm 0.51 \times 10^{-3}$ & 0.09 \\
11.10 & 11.30 & 11.24 & 9 & $1.24 \pm 0.41 \times 10^{-3}$ & 0.04 \\ \hline
\multicolumn{6}{c}{$0.6 < z < 0.9$} \\
11.10 & 11.40 & 11.25 & 9 & $1.17 \pm 0.47 \times 10^{-3}$ & 0.05\\
11.40 & 11.70 & 11.47 & 13 & $7.90 \pm 2.31 \times 10^{-4}$ &  0.03 \\
11.70 & 12.00 & 11.76 & 7 & $2.48 \pm 0.94 \times 10^{-4}$ & 0.07 \\ \hline
\multicolumn{6}{c}{$0.9 < z < 1.1$} \\
11.50 & 11.70 & 11.67 & 8 & $1.44 \pm 0.55 \times 10^{-3}$ & 0.06 \\ 
11.70 & 11.90 & 11.77 & 13 & $1.05 \pm 0.30 \times 10^{-3}$ & 0.07 \\ \hline
\end{tabular}
\end{minipage}
\label{lf_table}
\end{table}


\begin{figure}[hb]
\centering
\hspace{-0.5cm}
\includegraphics[width=8.5cm]{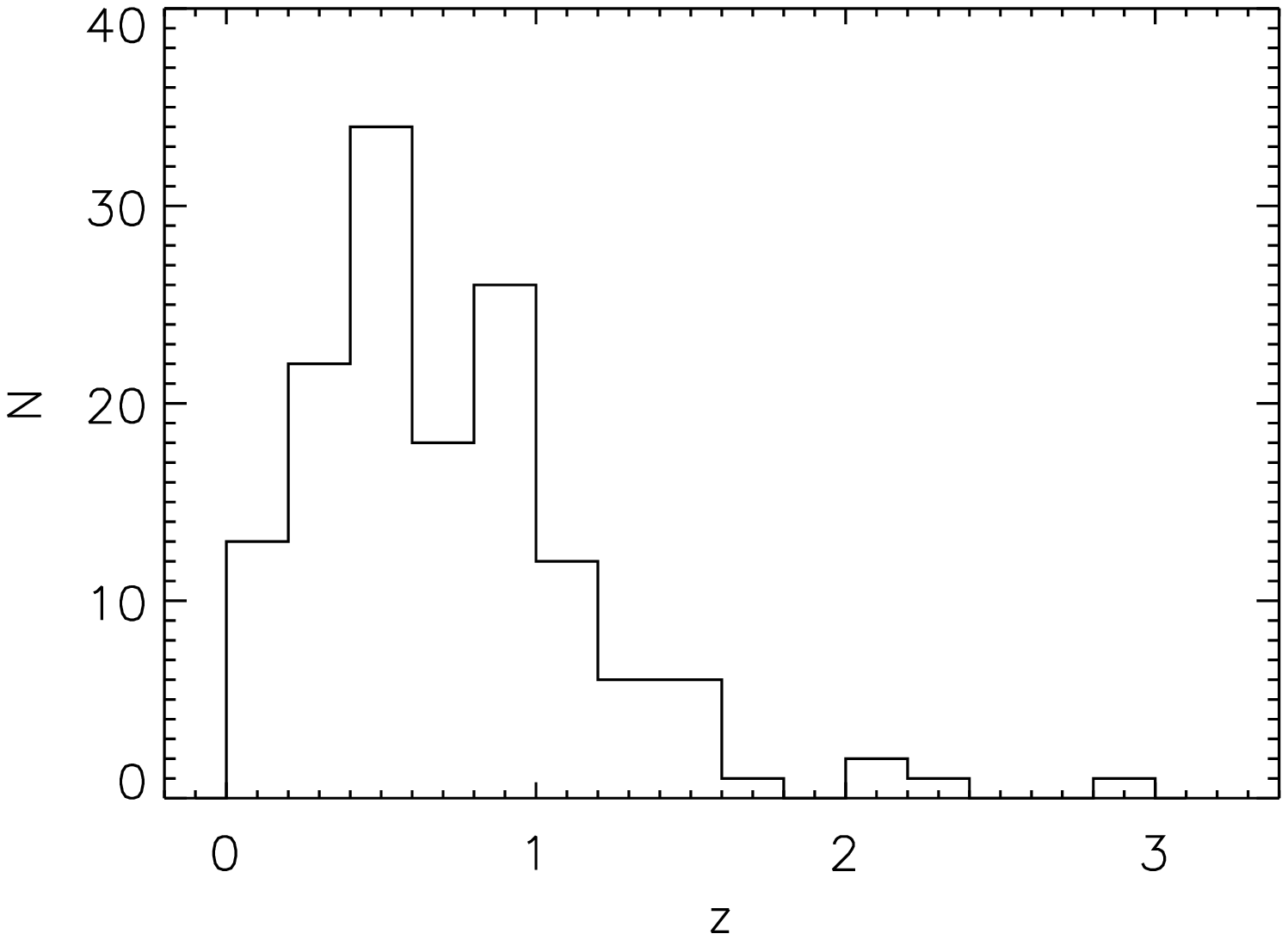}
\hspace{-0.5cm}
\includegraphics[width=8.5cm]{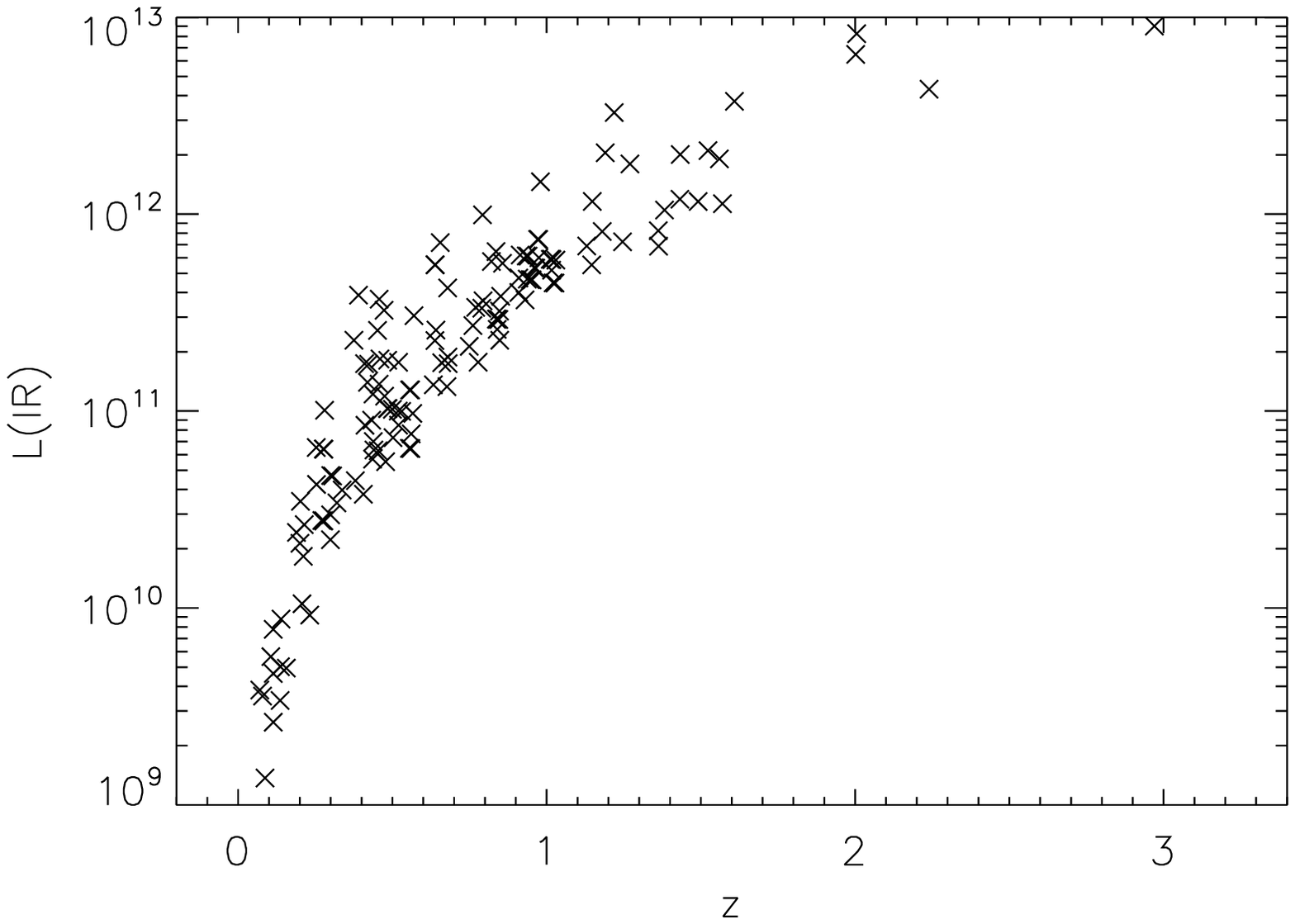}
\caption{{\em left}: The redshift distribution of the 70 $\mu$m sources. {\em right}: The IR luminosity (in $L_{\odot}$) versus redshift for the 70 $\mu$m sources.}
\end{figure}

\begin{figure}
\centering
\hspace{-0.5cm}
\includegraphics[width=14cm]{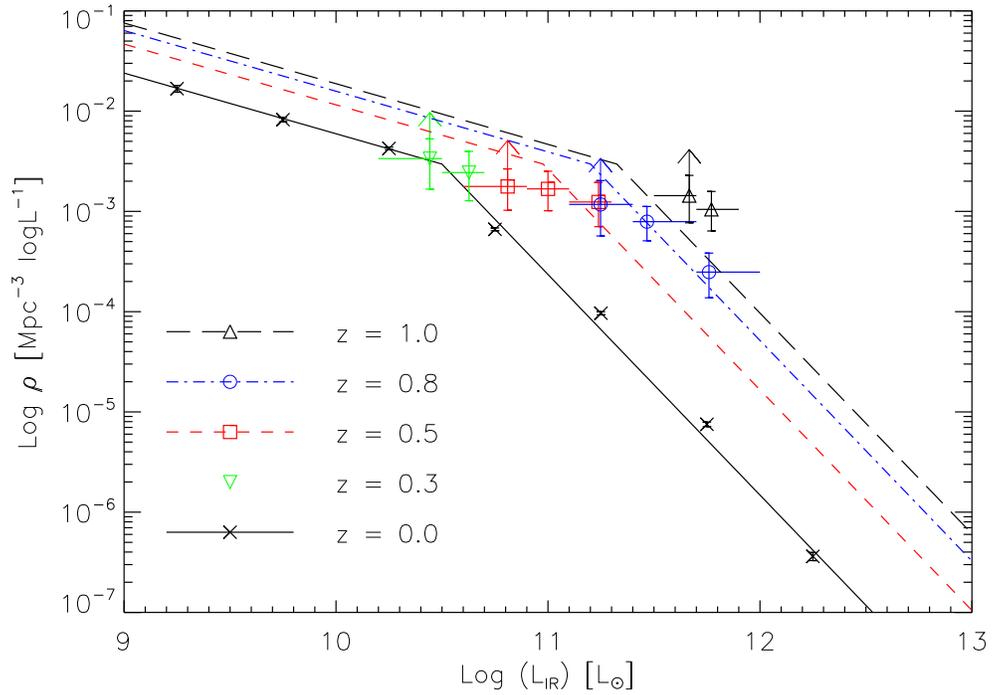}
\caption{The IR luminosity function (LF). Crosses mark the local LF from Sanders et al. (2003) and the corresponding solid line is the double power fit to the local data. The symbols mark the LF calculated in this work at redshift 0.3 (upside down triangles), 0.5 (squares), 0.8 (circles) and 1.0 (triangles). The lines are the local LF evolved to the corresponding redshift with the best fit pure evolution parameters. The arrows indicate bins which are incomplete because of the survey sensitivity. The horizontal error bars indicate the binsizes.}
\end{figure}

\end{document}